\RequirePackage[2020-02-02]{latexrelease}
\documentclass[english,aps,prper,reprint,showpacs,titlepage,longbibliography]{revtex4-2}   

\usepackage[T1]{fontenc}	
\usepackage[latin9]{inputenc}	
\usepackage{geometry}		
\usepackage{graphicx}
\usepackage[above,below]{placeins}	
\usepackage{times}
\usepackage{amsmath}

\usepackage{hyperref}  
\hypersetup{colorlinks=true,urlcolor=blue,citecolor=blue,linkcolor=blue}   
\usepackage{enumitem}          
\setlist{nosep}                 
\usepackage{cleveref}
\usepackage{censor}

\begin{document}

\begin{titlepage}

  \title{Using blogs to make peer-reviewed research more accessible}

  \author{Nicholas T. Young}
  \affiliation{Center for Academic Innovation, University of Michigan, 210 S. 5th Ave, Ann Arbor, MI 48104} 
  \author{Briley L. Lewis}
  \affiliation{Department of Physics and Astronomy, University of California Los Angeles 475 Portola Plaza, Los Angeles, CA, 90095} 
  \author{Emily Kerr}
  \affiliation{Department of Chemistry and Chemical Biology, Harvard University, 12 Oxford Street, Cambridge, MA, 02138}
  \author{Prasanth H. Nair}
  \affiliation{PERbites.org} 

  \begin{abstract}
    Discipline-based education researchers produce knowledge that aims to help instructors improve student learning and educational outcomes. Yet, the information produced may not even reach the educators it is intended to influence. Prior work has found that instructors often face barriers to implementing practices in peer-reviewed literature. Some of these barriers are related to accessing the knowledge in the first place such as difficulty finding and understanding research and a lack of time to do so. To lower these barriers, we created an online blog, PERbites, that summarizes recent discipline-based education research in short posts that use plain language. Having covered nearly 100 papers to date, we conducted a survey to see if we were addressing the need we had originally set out to address. We posted a 23-item survey on our website and received 24 usable responses. The results suggested that readers do generally agree that we are meeting our original goals. Readers reported that our articles were easier to understand and used more plain language than a typical discipline-based education research (DBER) journal article. At the same time, readers thought that all the important information was still included. Finally, readers said that this approach helped them keep up with DBER studies and read about papers they otherwise would not have. However, most readers did not indicate they changed their teaching and research practice as a result of reading our blog. Our results suggest that alternative methods of sharing research (e.g., non-peer reviewed publications or conference talks) can be an effective method of connecting research with practitioners, and future work should consider how we as a community might build on these efforts to ensure education research can make meaningful changes in the classroom.
 \clearpage
  \end{abstract}

  \maketitle
\end{titlepage}

\section{Introduction}\label{sec:intro}


Although science education research is continually investigating improvements to pedagogy and how students learn, there is often a disconnect between science educators at the K-12 and university levels and science education researchers. Ideally, educators would be able to access education research and apply its findings to their own teaching practice. In reality, however, educators often face a multitude of barriers in accessing education research. These can include journal paywalls or, in articles they can access, complex, jargon-filled language and methodology \citep{schaik_barriers_2018,shkedi_teachers_1998}. Teachers also often lack institutional support and working relationships with education researchers that would allow them to put education research into practice in their classrooms \citep{schaik_barriers_2018, shkedi_teachers_1998,henderson2008physics}. Even when faculty are aware of research-based teaching strategies, they often face further barriers in time, familiarity, and access when it comes to actually implementing them in the classroom \citep{henderson2007barriers,dancy2016faculty,dancy_pedagogical_2010}.

Yet, there are many opportunities to create greater connection between educators and education research, which can be informed by studies of instructors' behavior. Previous work has shown that instructors often access research in response to a particular classroom concern and that they tend to give more credence to research findings conveyed by a source they both trust and believe has practical knowledge of teaching \citep{drill2013teachers}. That study also found that instructors also expressed being more likely to read research findings if they were doing so as part of an initiative, committee, or other organized group or if the research was distilled into a shorter more focused format.

One such method to take advantage of these opportunities are blogs. Blogs can provide distilled research findings and help address the time and access barriers instructors face \citep{powell_using_2012}. Blog posts can be shorter, more focused, and less technical than primary research papers, and importantly are not paywalled. Blog efforts outside of education have been found to be successful in reaching community members who cannot access paywalled journals \citep{nelson_blogging_2019}. Blogs are also simple to set up, and can allow for two-way communication in comment sections, facilitating dialogue between researchers and practitioners. Additionally, plain language summaries of new research have been shown to be more effective at producing comprehension, a feeling of understanding, and enjoyment than the original abstracts do \citep{bredbenner2019video}.

One early attempt to use blogs to distribute science education research was the PERticles platform. Published on BlogPress from 2007 to 2009, this forum shared links and abstracts for physics education research \citep{hake2009over, noauthor_perticles_nodate}. In 2009, this moved to CiteULike; however, that platform shut down in 2019. 

More recently, PhysPort uses blogs as one medium to communicate about research-based teaching to physics faculty \citep{noauthor_physport_nodate}. These articles take the form of expert recommendations and answer common questions and concerns about teaching physics. They do not, however, share recent research directly.

These platforms left an opening in the space for blogs to distribute science education research, and to also focus on targeting both researchers and educators as the audience. In addition, no studies are published in the literature to record their efforts or analyze their readership or successes for future efforts to build on.

In this paper, we present a newer effort for a discipline-based education research (DBER) summary blog, PERbites. We will describe how we created PERbites and present the results of an initial study into its readership and effects in the community. Our study is guided by three questions:
(1) how can blogs about peer reviewed research help readers engage with the literature, (2) how can blogs about peer reviewed research help lower the barriers for engaging with the literature, and (3) how can blogs about peer reviewed research help readers learn about new teaching and research methods and apply them in their work?



\section{About PERbites}\label{sec:about}

PERbites is a graduate-student organization that publishes an online DBER literature blog \citep{noauthor_perbites_nodate}. PERbites grew out of the larger ScienceBites project, a collection of online blogs covering various scientific disciplines, with the goal of making research in their fields more accessible \citep{noauthor_sciencebites_nodate}. One of the authors identified the potential for a similar site to be useful in DBER. The purpose of the site is to make current DBER research accessible for early career DBER researchers, as well as physics education practitioners of all career stages and other interested groups. Our hope is to positively impact the recognized challenges of educators accessing and utilizing DBER, as described in Section \ref{sec:intro}. Specifically, PERbites's three main goals are to (1) help researchers and educators keep up with recent DBER work; (2) lower barriers around understanding DBER papers; (3) expose readers to new teaching and research methods, so that they can apply them in their work. These goals are broadly aligned with common aims of plain language summaries according a recent review \citep{stoll_plain_2021} and what readers of evidence summaries and blog posts care about \citep{barbara2016user}. Based on the authors' estimation, around 40\% of the posts to date advance goal 3 by exposing readers to new teaching and research methods, while the remaining posts cover `pure research', advancing the goal of helping researchers and educators keep up with DBER.

Beginning in 2018, PERbites has published summaries of nearly 100 peer-reviewed papers. Our articles are written by early career researchers with backgrounds or interests in STEM education and/or science communication. Each article takes around 5-6 hours to prepare and write. After publication, each article is emailed to our subscribers and advertised on our social media platforms. Select posts are also shared by the larger ScienceBites group on social media and have been featured in an American Association of Physics Teachers newsletter, an American Physical Society forum, and multiple podcasts. To establish itself as a long-standing pillar of the community that can be relied on for updates on relevant education research, PERbites posts between 1-2 article summaries each month and has done so the past 4 years. This approach is supported by research that has found that successful dissemination approaches typically need to engage participants over time to be successful; one-time workshops or a collection of minimally related workshops have not been found to be effective \citep{henderson_facilitating_2011}.

Each blog post follows a standard model, known as the ``bites'' format, 
to create an accessible summary of a recent, interesting paper. Additionally, each PERbites post should be able to be read in approximately 5 minutes, addressing the barrier that many academics and educators do not have enough time to keep up with recent literature. Each of our articles follows a standard format as detailed below:
\begin{itemize}
    \item Clearly state the article title, authors, their affiliations, and where to access the full text.
    \item Describe the overarching problem and why it matters.
    \item Explain how the researchers have made progress on this problem with this new research.
    \item Conclude with a re-statement of what the new takeaway is, as well as a recap of what we still do not know. (Note: Communicating uncertainty can positively influence readers' perception of trustworthiness of results in the long-term and contribute to a more accurate conception of scientific methodology. \citep{frewer2002public,retzbach2015communicating,kreps2020model,van2020effects}. In addition, "overly rosy" descriptions of how well a research-based instructional strategy works is associated with instructors discontinuing the strategy \citep{henderson_use_2012}.)
    \item Briefly explain the implications of this conclusion.
\end{itemize}

Unlike other disciplines, there is a distinct challenge in covering DBER in that there is also a large population of non-DBER researchers interested in DBER results (e.g., physics faculty, high school physics teachers, physics and astronomy graduate student teaching assistants, etc.). Other ScienceBites sites are aimed at helping undergraduate students and early career researchers in a field gain exposure to literature in an understandable format. Although early career education researchers can use PERbites for the same reason, PERbites is unique among the ScienceBites sites in that it aims to serve this additional population of education practitioners outside the field of DBER research. 

Other ScienceBites efforts, such as Astrobites, have conducted readership surveys to better understand their audience and educational studies to explore their effectiveness and classroom applications \citep{sanders_incorporating_2017,sanders2012preparing,2019BAAS...51g.230K}. However, given the unique nature of PERbites's audience and purpose among the ScienceBites sites, this effort warrants dedicated study.

\section{Methods}\label{sec:methods}
In early 2022, we posted a 23-item survey on our website asking readers of PERbites about how successful they thought we were in achieving our three goals. All ($\sim$100) subscribers to PERbites received an email announcing the survey and we advertised the survey on our Facebook and Twitter accounts. The survey included Likert-scale questions for assessing the impact of PERbites, multiple choice questions around participants' usage of PERbites and their professional backgrounds, and free response questions for their demographics and any additional information they wished to provide. Likert-scale questions consisted of 5 options on the scale of strongly disagree to strongly agree and included a N/A option as well. We used Qualtrics \citep{noauthor_qualtrics_2005} to collect the data and a copy of the survey can be found at \citep{perbites_survey_nodate}. The survey items were designed for this effort, and do not come from a pre-validated survey. The IRB at the University of Michigan ruled this study exempt (HUM00211844).

Of the 33 people who agreed to participate, 24 completed the survey. We will only be examining those responses.

Given the small sample size, we will only report raw numbers rather than conduct any statistical analyses. Under this approach, we acknowledge that our results are not intended to be generalizable,  but rather, our results serve to document an example of a discipline-based education research dissemination effort. We believe such efforts should be documented so that the community has record of what efforts have been tried, which efforts show potential, and which do not.

\section{Results}\label{sec:results}


\begin{figure*}
  \centering
  \includegraphics[width=.75\linewidth]{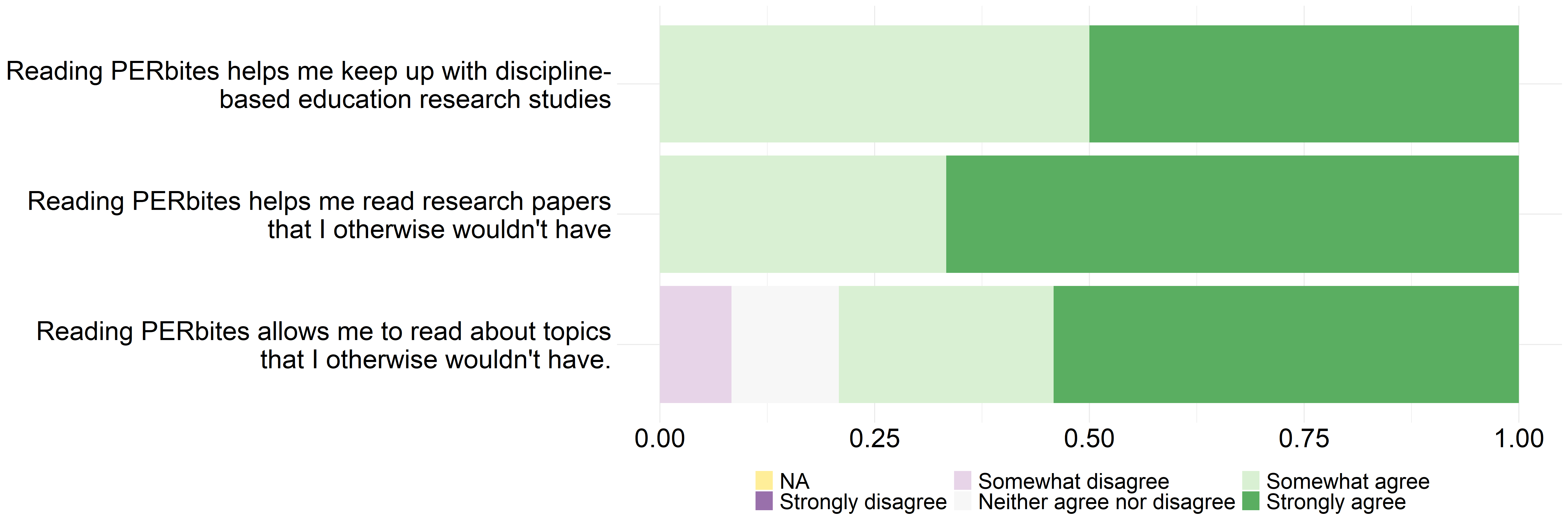}
  \caption{Distribution of responses to the questions about the success of our goals around dissemination of discipline-based education research. For all of the questions, most respondents answered agree or strongly agree.}
  \label{fig:Q5s}
\end{figure*}

\begin{figure*}
  \centering
  \includegraphics[width=.75\linewidth]{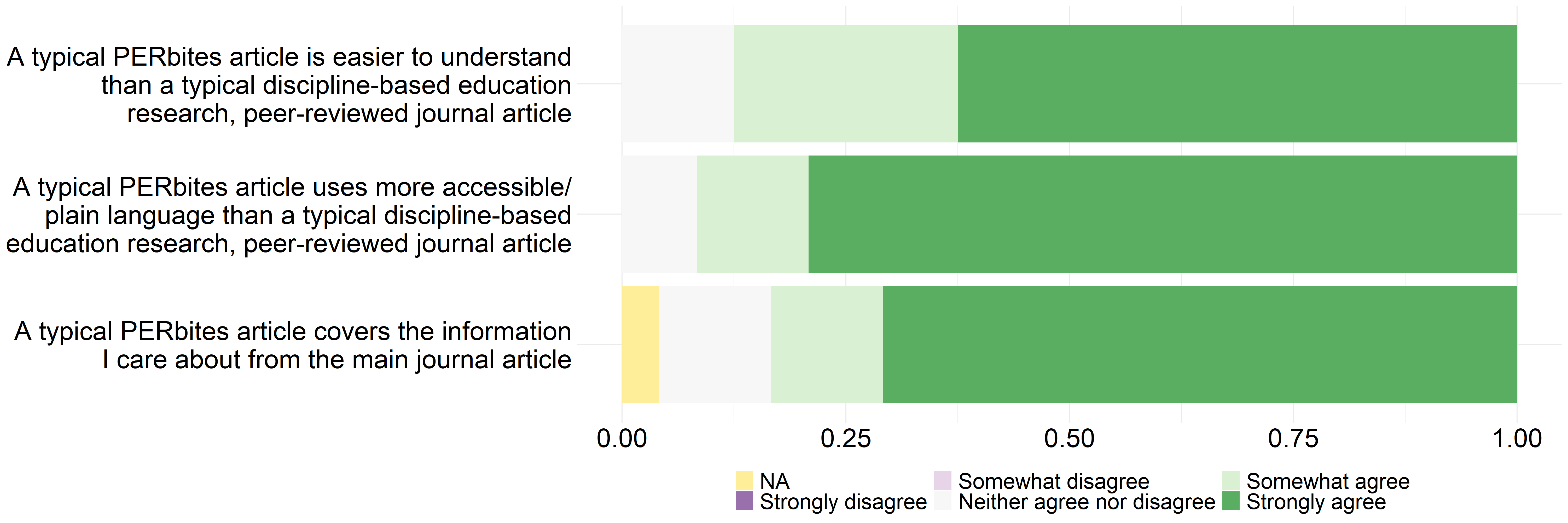}
  \caption{Distribution of responses to the questions about the success of our goals around lowering barriers to understanding or reading peer-reviewed literature. For all of the questions, most respondents strongly agreed that we were successful in our goals.}
  \label{fig:Q7s}
\end{figure*}

\begin{figure*}
  \centering
  \includegraphics[width=.75\linewidth]{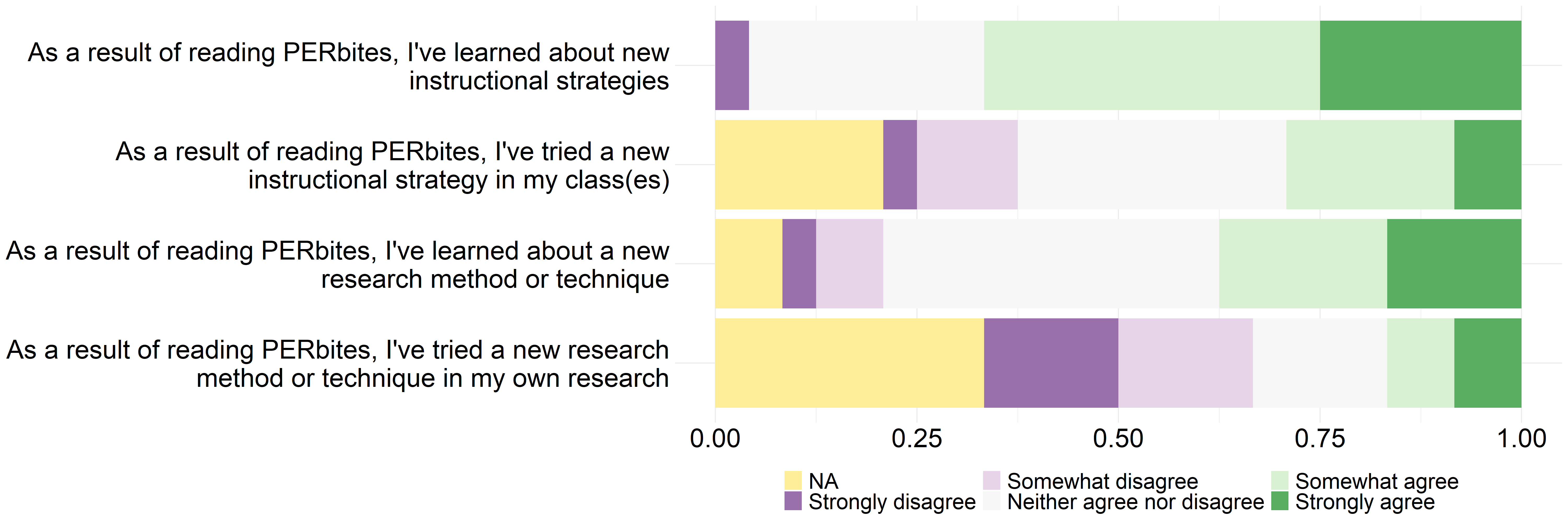}
  \caption{Distribution of responses to the questions about the success of our goals around teaching and research. Here, we were less successful in our goals. While many respondents reported learning about new techniques, only a third reported trying them.}
  \label{fig:Q6s}
\end{figure*}



Before looking at the impact of PERbites, it is useful to understand who participated in the survey. When we look at the occupations and disciplines, we find that 8 reported four-year college or university instructors (6 in physics and 2 in astronomy), 3 reported astronomy graduate students, 5 reported high school teachers (2 in physics, 1 in education, 1 in geoscience, and 1 in an undisclosed discipline), 2 reported postdoctoral researchers (1 in physics and 1 in astronomy), 1 reported physics researcher at a four-year college or university, 1 reported physics staff, 1 reported physics instructor at a two-year college, and 3 did not report anything (1 in astronomy, 1 in engineering, and 1 in an unknown discipline).

Regarding gender, we find that 9 respondents identified as male, 9 identified as female, 1 identified as woman, 1 identified as female-adjacent, 1 identified as a GNC [gender non-conforming] woman, 1 identified as NB [nonbinary], and 2 respondents did not disclose their gender. 

When looking at race, we find that 19 respondents identified as white, 2 identified as Caucasian, 1 identified as white, non-Hispanic, and 2 respondents did not disclose their race. As all of the participants who disclosed their race are white, we did not look at any intersectional identities. This may be indicative of the general demographics of physics as a discipline \citep{american} and signals a need for PERbites to consciously expand its reach in the future.

While not included on the survey, visitor statistics collected from our website provide us additional insight into readers of PERbites. Since launch, our site has received over 30,000 page views. Country-level data suggest that nearly 66\% of our site views come from users based in the United States. After the United States, the largest percentage of readers are based in the United Kingdom, Canada, and India.



Looking across the results of the survey questions (\Cref{fig:Q5s,fig:Q7s,fig:Q6s}), we find that readers overwhelmingly agree that PERbites is successful in helping them keep up with discipline-based education research and lowering barriers around understanding papers. However, we find limited evidence that reading PERbites changes respondent's practices.

Figure \ref{fig:Q5s} shows participants' responses to our three questions about disseminating discipline-based education research. For the first two questions, all participants agreed or strongly agreed that PERbites helps them keep up with DBER studies and that PERbites allows them to read about papers they otherwise would not have. In the case of the third question, 5 of the 24 participants did not agree to any degree that PERbites allows them to read about topics they otherwise would not have. 

We find a similar story for the questions presented in Figure \ref{fig:Q7s}, which covers our goals around lowering barriers to understanding or reading peer-reviewed literature. For all three questions, we find that a majority of participants strongly agreed that PERbites is easier to understand than a typical DBER publication, uses more accessible language than a typical DBER publication, and at the same time, still includes the information that readers care about.

In contrast, the questions presented in Figure \ref{fig:Q6s} tell a different story. While most participants tended to agree that PERbites is successful in introducing readers to new instructional strategies, that was not the case for introducing readers to new research techniques or affecting instructional or research practice. In the case of trying a new instructional strategy as a result of reading PERbites, only 7 of the 24 participants agreed to some degree that they have tried something they learned as a result of reading PERbites. In terms of trying something new in research, only 4 of the 24 participants agreed to some degree. In both cases however, there are a relatively large amount of "Not Applicable" answers, likely the result of instructors not engaging in research as part of their position and researchers not engaging in instruction as part of their position.

\section{Discussion}\label{sec:disc}
Returning to our research questions, we find that blogs can help readers engage with DBER literature and lower the barriers for engaging with it. Our reader survey indicated that our articles about peer-reviewed research are easier to understand and use less jargon, addressing two of the barriers prior work has found educators encounter when trying to engage with peer-reviewed studies \citep{schaik_barriers_2018,shkedi_teachers_1998}. We acknowledge that this is likely a result of the different audiences served by our blog and peer-reviewed literature. In the case of peer-reviewed literature, the intended audience is often narrowly defined to be other researchers in the sub-field of the research. In the case of our blog, the audience is broadened to include DBER researchers and STEM educators. Due to this change in audience, the writing style must also adapt.

In terms of helping readers engage with the literature, all participants reported reading about papers they otherwise would not have. When it came to reading about topics they otherwise would not have, a minority of survey respondents reported that that was not the case. As many of the participants are educators, it would not be surprising if they read about some of the same topics PERbites covers elsewhere, resulting in a non-agree response to this question.

When considering how reading our blog affected actual practice, the results were less promising. While most survey respondents said they learned a new instructional strategy, few reported trying it in their classrooms. Even fewer reported learning about a new research technique or trying it in their work.

While the results in this area are discouraging, they are not unexpected. Prior work on physics faculty implementing research-based instructional strategies in their courses has found that situational barriers, not dissemination efforts, are the key problem for adoption of new practices \citep{henderson2007barriers}. Given that our blog does not support faculty in making changes in their classroom or address situational barriers at their institution, we should not expect that simply introducing faculty to new instructional ideas would automatically lead to the adoption those strategies in their classrooms in most cases.

\section{Conclusions and Future Work}\label{sec:conc}
Our work suggests that blogs can be successful in reducing barriers for accessing and understanding peer-reviewed research, and their usefulness as a dissemination method should be explored further. It also opens a broader conversation about how the results of DBER can be shared to reach educators. Such conversations should include who the responsibility for disseminating research falls upon. Under our approach, a group of early career researchers has been responsible for sharing other people's research. Yet, paper authors themselves can adapt our approach, as some in the medical community have called for \citep{gudi2021plain}. Alternative approaches to research dissemination, such as sharing graphical abstracts, video abstracts, and Twitter threads and discussions, are also gaining traction in education research (e.g., \citep{rodrigues2021get, doucette2020hermione}), but are restricted to a limited number of journals and researchers. Future work should then consider how to integrate these dissemination methods into new and existing DBER publication and award structures (e.g., tenure and promotion, professional society awards, etc.). Additionally, future work should conduct efficacy studies to determine how best to use these tools and how they might be improved. At the same time, future work should examine what barriers researchers face in adopting these dissemination methods and what resources and supports they would need to be successful.

Given that these efforts are generally run by early career researchers, future work could also consider how participating in such dissemination efforts affects those doing this research dissemination work. For example, does participating in such efforts improve the participants' communication skills, allow them to explore new career interests, or provide some other benefit? While such studies are not new (e.g., \citep{hinko2014use, fracchiolla2016university, prefontaine2018intense, fracchiolla2020participation, Rethman2021impact}), focusing on additional programs with different audiences can provide a more complete picture of the benefits participants might receive.

Taken together, conducting studies of dissemination efforts (such as this work) can inform future efforts so that they have positive impacts on both on instructors and the individuals who participate in such efforts.

\acknowledgments{
We would like to thank the American Astronomical Society for their support and hosting of PERbites and the ScienceBites collaboration for their support in launching PERbites.
}

\bibliography{mybibfile} 

\end{document}